\documentclass[aps,prb,twocolumn,superscriptaddress,longbibliography]{revtex4-1}
\usepackage{graphicx}
\usepackage{amsmath}
\usepackage[colorlinks=true,allcolors = blue]{hyperref}
\usepackage{xcolor}

\begin{document}
	
	
	\title{Strongly enhanced temperature dependence of the chemical potential in FeSe}
	
	
	\author{L. C. Rhodes}
	\affiliation{Department of Physics, Royal Holloway, University of London, Egham, Surrey, TW20 0EX, UK}
	\affiliation{Diamond Light Source, Harwell Campus, Didcot, OX11 0DE, UK}
	\author{M. D. Watson}
	\affiliation{Diamond Light Source, Harwell Campus, Didcot, OX11 0DE, UK}
	\author{A. A. Haghighirad}
	\affiliation{Clarendon Laboratory, Department of Physics, University of Oxford, Parks Road, Oxford OX1 3PU, UK}
	\author{M. Eschrig}
	\affiliation{Department of Physics, Royal Holloway, University of London, Egham, Surrey, TW20 0EX, UK}
	\author{T. K. Kim}
	\affiliation{Diamond Light Source, Harwell Campus, Didcot, OX11 0DE, UK}
	
	
	
	\date{\today}
	\begin{abstract}
		Employing a 10-orbital tight binding model, we present a new set of hopping parameters fitted directly to our latest high resolution angle resolved photoemission spectroscopy (ARPES) data for the high temperature tetragonal phase of FeSe. Using these parameters we predict a large 10~meV shift of the chemical potential as a function of temperature. 
		In order to confirm this large temperature dependence we performed ARPES experiments on FeSe and observed a $\sim$25~meV rigid shift to the chemical potential between 100~K and 300~K.  
		This unexpectedly strong shift has important implications for theoretical models of superconductivity and of nematic order in FeSe materials. 
	\end{abstract}
	
	
	\maketitle
	

	\section{Introduction}
	To understand high-temperature superconductivity and nematic order in iron-based superconductors, it is necessary to obtain an accurate description and understanding of their electronic structure. 
	However, modelling the electronic structure of iron-based superconductors has proved to be a challenging task. 
	Ab-initio calculations such as density functional theory (DFT) show some disagreement with quantum oscillation experiments \cite{Coldea2008}, as well as with the band dispersions obtained from ARPES \cite{Ding2009,Nakayama2009}. In fact, the calculated models need to be renormalized by a factor between 2-10 to be in qualitative agreement with experimental results \cite{Yin2011}. 
	More sophisticated theoretical treatments such as dynamical mean field theory (DMFT) are able to account for the orbital-dependent band renormalizations \cite{Aichhorn2010}, yet they do not account for more specific features, such as the shrinking of the hole and electron pockets seen in even the simplest iron-based superconductor, FeSe \cite{Watson2016_2}.
	
	FeSe has attracted a lot of interest due to the highly tuneable nature of the superconducting transition temperature; ranging from 8~K in bulk samples \cite{Hsu2008}, 37~K under pressure \cite{Medvedev2009}, 43~K with intercalation \cite{Burrard-Lucas2013} and 65~K with a monolayer grown on SrTiO$_3$ \cite{He2013}.  
	At 90~K bulk FeSe also experiences a nematic transition - a form of electron correlation which breaks four-fold rotational symmetry\cite{Fernandes2014}, yet unlike most other iron based superconductors, it is not accompanied by an antiferromagnetic phase transition.
	In recent years, ARPES studies on FeSe have yielded increasingly detailed information on the quasiparticle dispersions \cite{Borisenko2016,Watson2015,Watson2016,Federov2016,Fanfarillo2016}, yet despite the theoretical interest, a fully accurate model consistent with experiment is still lacking. 
	
	In this paper we present a new set of tight binding parameters fitted directly to experimental data for the high temperature tetragonal phase of FeSe.
	Our parameters are optimized against high-resolution ARPES data at 100~K and provides a quantitatively accurate description of the observed 3D electronic structure. Using this parameter set we predict a large 10~meV rigid increase of the chemical potential between 100~K and 300~K, not predicted from a DFT-based parameter set. Motivated by this finding, we have also performed a temperature-dependent ARPES study on FeSe up to 300~K and observed an enhanced chemical potential shift of $\sim$ 25~meV. We discuss a possible explanation for this enhancement of the observed chemical potential shift with respect to the model prediction. Our results highlight the crucial importance of complete consideration of the chemical potential in both theory and experiment.

	\section{Experimental Methods}
	High-quality single crystal samples of FeSe were grown by chemical vapor transport \cite{Bohmer2013}. ARPES measurements were performed at the I05 beamline at the Diamond Light Source, U.K. \cite{Hoesch2016}. The Fermi level position was calibrated by fitting the Fermi function to freshly deposited polycrystalline gold at 35~K. All maps and band dispersions presented here were measured with a photon energy of 56~eV. This ensured that only a single reference measurement from the polycrystalline gold was needed to calibrate the binding energy, eliminating a possible source of experimental error. 
	
	Band dispersions used for the optimization of the tight binding model were taken from data published in Ref.~[\onlinecite{Watson2015}] and [\onlinecite{Watson2016}] by fitting a Lorentzian to the momentum distribution curves at 100~K for photon energies of 37~eV ($k_z = 0$) for the $\Gamma$ and $M$ points and 56~eV ($k_z = \pi/c$) for the $Z$ and $A$ points.
	
	\section{Tight Binding Model}
	We use the 10-orbital tight binding model of Ref.~[\onlinecite{Eschrig2009}] with the inclusion of an additional spin-orbit coupling term \cite{Saito2015} as the framework for our phenomenological optimization. 
	This model includes all 10 $d$-orbitals associated with the $P4/nmm$ crystallographic space group, i.e the 2-Fe unit cell. 
	Previous tight binding models of iron based superconductors  \cite{Graser2010,Knolle2012,Mukherjee2015} have normally made use of the glide symmetry of the 2-Fe unit cell to unfold the Brillioun zone into a pseudo $\mathbf{k}$-space corresponding to a 1-Fe unit cell \cite{Anderson2011}.
	Whilst using the 1-Fe unit cell has been shown to be a good approximation \cite{Mukherjee2015}, and is exact in the absence of spin-orbit coupling \cite{Nica2015}, optimization of our model against experimental ARPES data allowing for spin-orbit coupling requires a 2-Fe unit cell treatment. 
	
	In Fig. 1 we show the calculated band structure obtained using our phenomenological tight binding parameters fitted to ARPES data obtained in the high temperature tetragonal phase at 100~K . We present our ARPES-based parameters in the appendix. These parameters were optimised at both $k_z = 0$ ($\Gamma$ and M) and $k_z=\pi$ (Z and A) providing a complete three dimensional description of the electron structure of FeSe.
	The addition of the spin-orbit interaction term manifests itself in two ways. First, it lifts the degeneracy of the $d_{xz}$  and $d_{yz}$ bands revealing a 20~meV splitting at $\Gamma$ and Z  \cite{Borisenko2016} and secondly, it creates an anticrossing of the $d_{xy}$ and $d_{xz}$ near the M and A points. 
	The parameters also correctly describe the recently observed separation at the M and A points \cite{Watson2016,Federov2016} between the degenerate $d_{xz}$ /$d_{yz}$ band at -22~meV and the $d_{xy}$ band at -42~meV. In the ARPES data there is also a $d_{z^2}$ band at $\sim$~-220~meV, which is accounted for in the model, although not shown.

	In this tight binding model, the size of the electron pocket and flatness of the $d_{xy}$ band are controlled by the same hopping parameters. Since we wish to maintain quantitative agreement at the Fermi surface, it became necessary to have an exaggerated $d_{xy}$ dispersion at higher energies. Nevertheless, the binding energies of the bands at the high symmetry points are in agreement with experiment, as are the dispersion of all the bands close to the chemical potential. 
	Consequently, as the parameters were fitted within the region of +10~meV to -100~meV we can not comment on the reliability of the model at higher energies.
	
	This highly accurate description of the low-energy part of the band dispersions near the chemical potential allows us to quantitatively reproduce the experimentally determined Fermi surface. Our model therefore provides a good starting point for discussions of low-energy properties of FeSe.  
	
	\begin{figure}[t]
		\includegraphics[width = 0.5\textwidth]{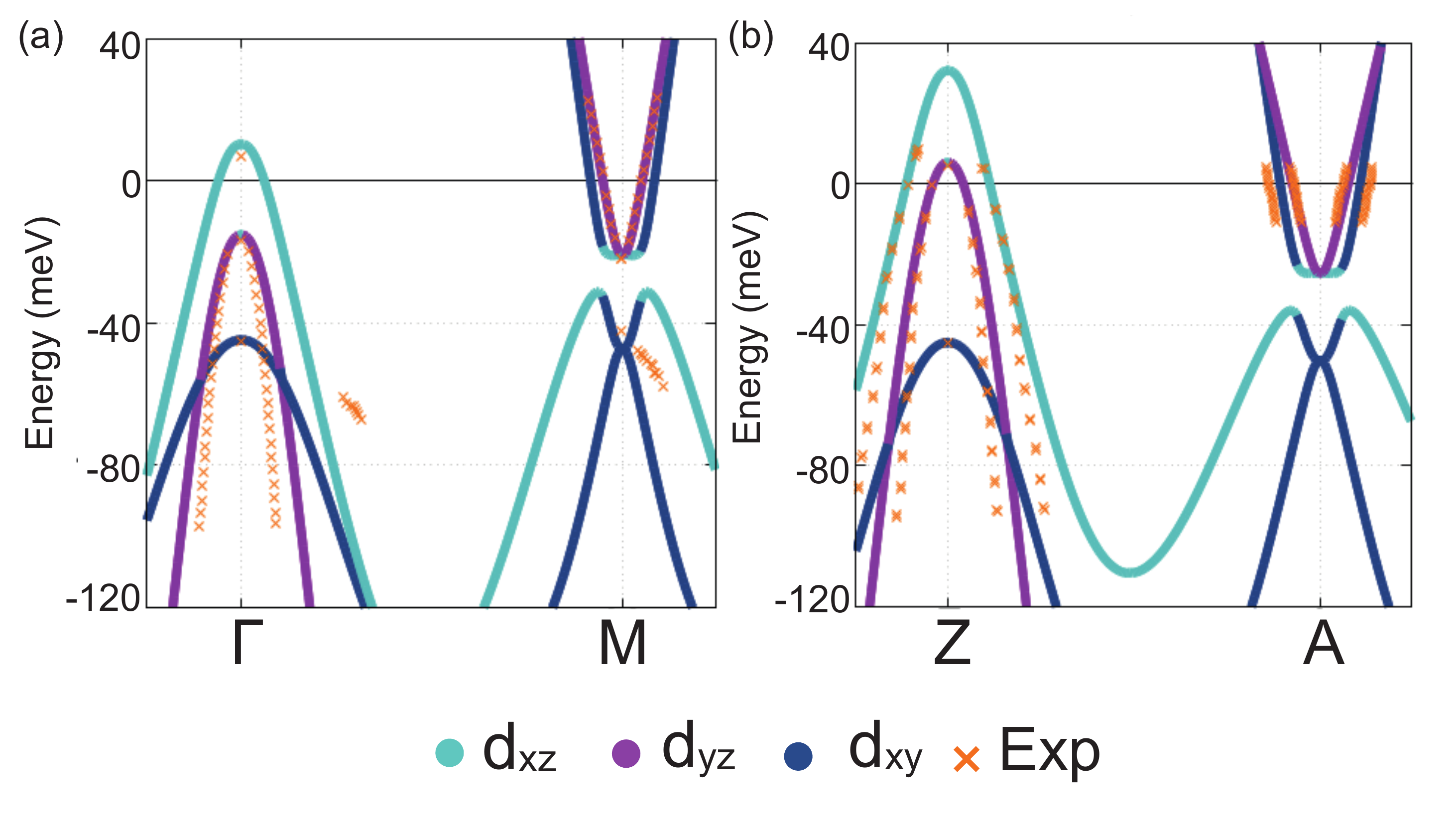}
		\caption{\label{figure1} Tight binding band dispersion calculated using our ARPES-based parameters, shown for $k_z=0$ (a) and $k_z=\pi$ (b), together with experimental data points used to constrain the model (orange crosses).
		}
	\end{figure}

	\section{Chemical potential}
	Next we consider what turns out to be an important issue in FeSe, the temperature dependence of the chemical potential.
	Temperature dependent ARPES studies on the iron-based superconductors BaFe$_2$As$_2$ \cite{Brouet2013}, 
	Ba(Fe$_{1-x}$Ru$_x$)$_2$As$_2$ \cite{Dhaka2013} 
	and FeSe$_{1-x}$S$_x$ \cite{Abdel-Hafiez2016}, as well as the semi-metal WTe$_2$ \cite{Wu2015}
	have all observed a substantial change to the $k_F$ values of the hole and electron pockets, consistent with a rigid shift to the chemical potential as a function of temperature. This can be explained by the shallow nature of the hole and electron pockets in the iron-based superconductors. To maintain the total charge of the system, the chemical potential must shift to counterbalance any asymmetry in the density of electron and hole carriers as a function of an external variable such as temperature. This effect is small if the density of states around the chemical potential is uniform for both the hole and electron bands, such as in an unrenormalized DFT-based model. However for the shallow band dispersions observed experimentally, with the top of the hole and bottom of the electron bands within a few $k_BT$ of the chemical potential, a change in temperature can have an anisotropic effect on the density of thermally active electron and hole carriers, necessitating a shift of the chemical potential.
	
	\subsection{Theoretical predictions }
	To determine the temperature evolution of the chemical potential in FeSe we calculate the total number of electrons, $N$,
	\begin{equation}
		\label{particleNum}
		N =2\sum_{\mathbf{k}}\sum_\nu f(E_\nu(\mathbf{k})-\mu),
	\end{equation}
	\noindent where $\nu$ is the band index, 
	$E_\nu(\mathbf{k})$ is the energy of band $\nu$ at momentum $\mathbf{k}$,
	$f(E_\nu(\mathbf{k})-\mu) = [1+e^{\beta(E_\nu(\mathbf{k})-\mu)}]^{-1}$ is the Fermi function with $\beta = \frac{1}{k_{\rm B}T}$, and  
	$\mu$ is the chemical potential.
	At 100~K the particle number was determined to be 12.00$e^-$, for a $\mathbf{k}$-grid of size 64x64x8. This is consistent with the total number of valence electrons available for two Fe$^{2+}$ atoms in a 2-Fe unit cell. We fix $N$ and calculate $\mu(T)$ from Eq.~\eqref{particleNum} numerically. 
	
	\begin{figure}[t]
		\includegraphics[width = 0.5\textwidth]{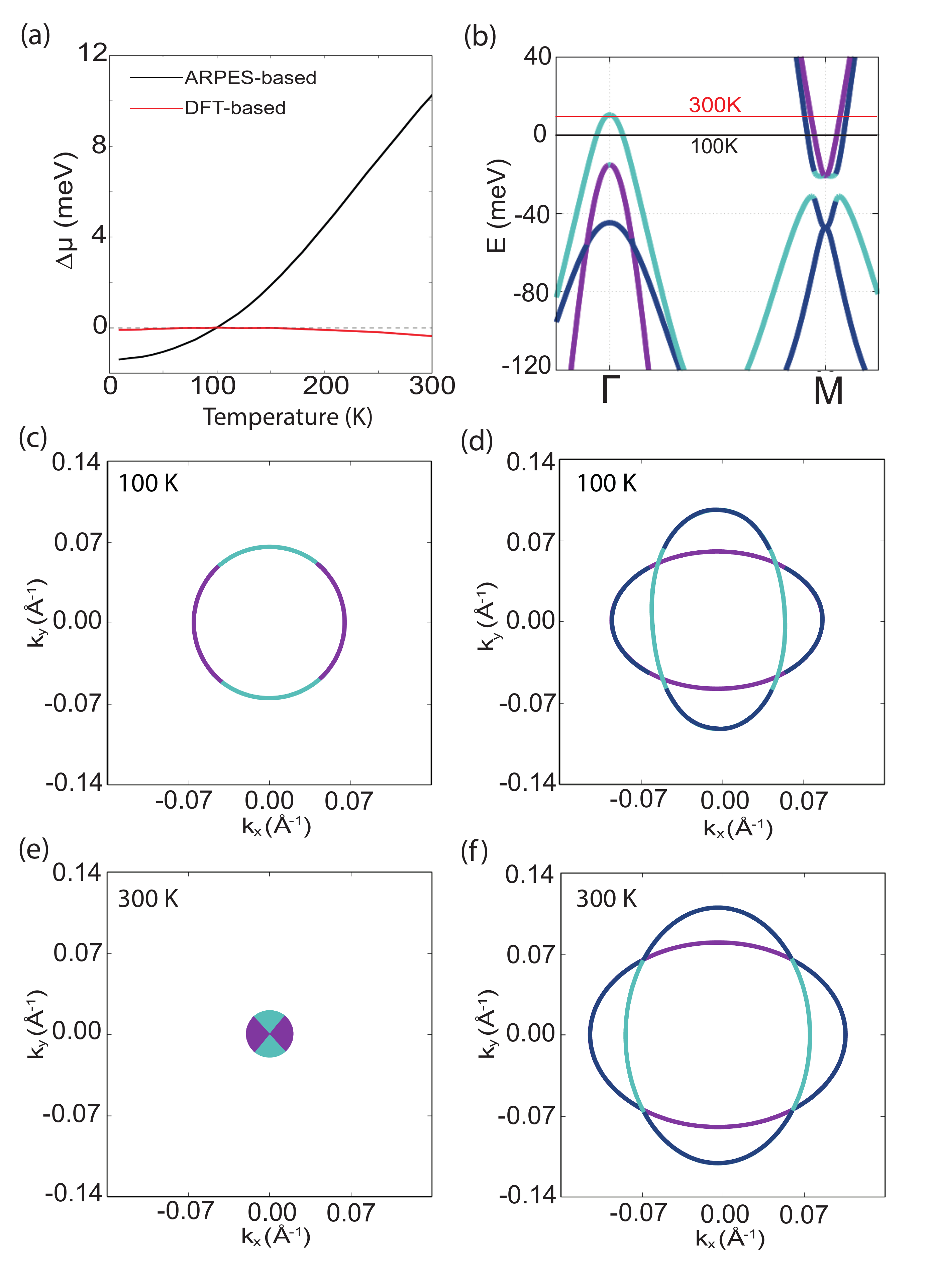}
		\caption{\label{figure2} Predicted temperature effects of the chemical potential in FeSe. (a) Chemical potential as a function of temperature predicted using our ARPES-based parameters (black line), compared to a set of DFT-based parameters (red line). b) Band dispersions along $\Gamma-M$ indicating the location of the chemical potential at 100~K and 300~K. c-f) The corresponding Fermi surface maps for the hole (left) and electron (right) pocket at 100~K (c,d) and 300~K (e,f). }
	\end{figure}

	Fig.~2(a) shows the results for the evolution of the chemical potential up to 300~K as predicted from the ARPES-based parameters. This is compared to the predicted evolution of the chemical potential using DFT-based parameters taken from Ref. [\onlinecite{Eschrig2009}]. Whilst the DFT-based parameters would give rise to only a very small change of -0.4~meV between 100~K and 300~K, the ARPES-based parameters we find a much larger 10~meV increase to the chemical potential.
	
	In the case of the DFT-based parameters, the unrenormalized hole and electron bands crossing the chemical potential appear as large pockets with a wide bandwidth and relatively uniform density of states close to the chemical potential. Hence, the chemical potential hardly changes as a function of temperature as would normally be the case in a metallic system. However for the model based on the experimental band dispersions, much smaller Fermi surfaces and renormalized band dispersions close to the chemical potential are described, including a very shallow top of the hole band. Thus the thermal population of hole and electron states can become rather asymmetric, leading to a much more significant shift of the chemical potential. As a result within our realistic ARPES-based parameter set, we find a non-trivial chemical potential effect, particularly for temperatures exceeding 100 K.
	
	As the top of the hole band at $\Gamma$ is $\sim$10~meV above the chemical potential, a shift as large as we predict would cause a substantial change to the Fermi surface topology. In Fig.~2(b) we show where the chemical potential would cut the band structure at 100~K and 300~K. By 300~K the chemical potential would just be touching the top of hole band at $\Gamma$. The effect of this on the Fermi surface is shown in Fig.~2(c-f). The chemical potential increase would lead to a dramatic reduction of the $k_F$ value of the hole pocket as a function of temperature, and an increase to the $k_F$ value of the electron pocket.

	\begin{figure*}
		\includegraphics[width = 0.98\textwidth]{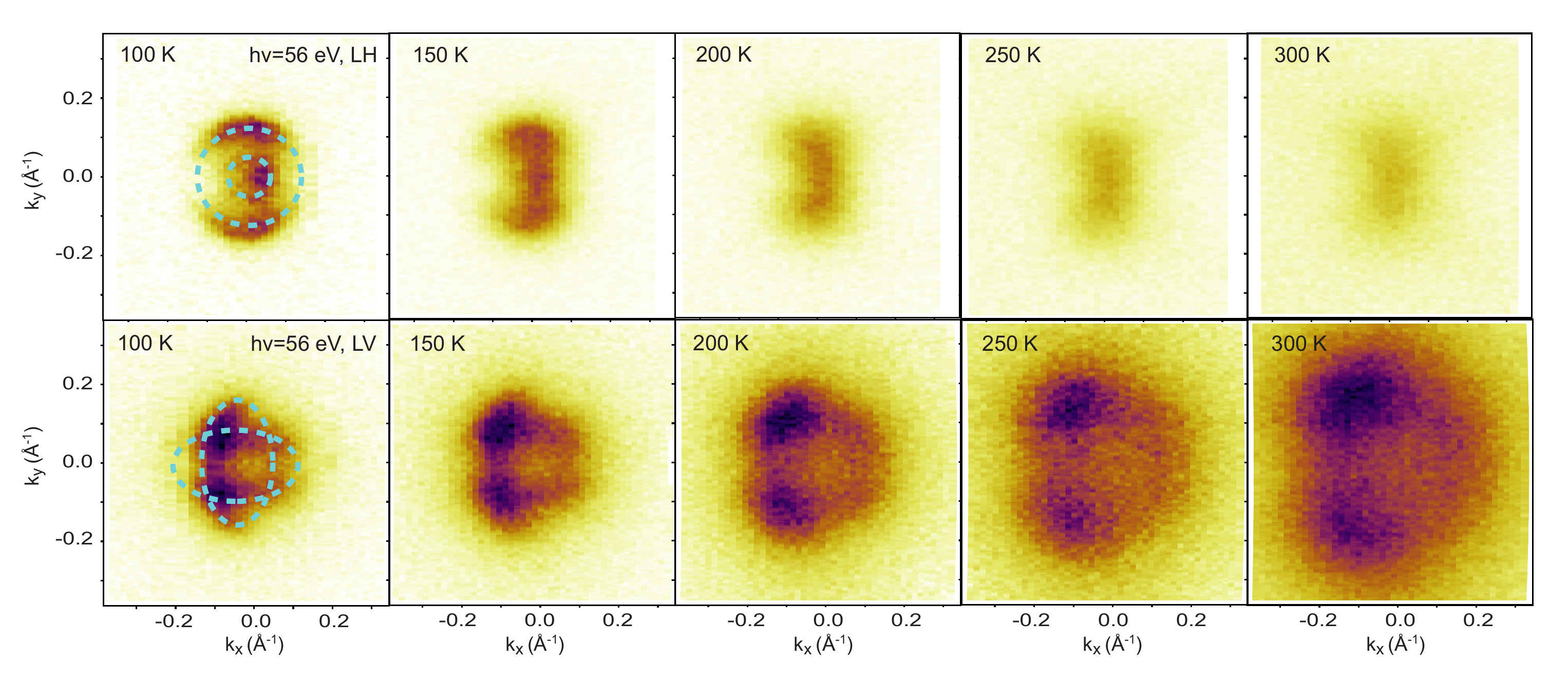}
		\caption{\label{figure3} Temperature dependent Fermi surfaces of the hole pocket (top row) and electron pocket (bottom row), taken at a photon energy of 56~eV between 100~K and 300~K. ARPES intensity was integrated within a $\pm5$~meV range around the Fermi level. The axes are defined relative to the center of the pocket.}
	\end{figure*}

	\subsection{ARPES results}
	To confirm the predicted temperature dependence of the chemical potential we tracked the evolution of the hole and electron pockets at the high symmetry Z and A points by ARPES.

	In Fig.~3 we show the temperature dependence of the hole (Z point) and electron (A point) pockets as a function of temperature. At 100~K the hole pocket has two circular bands crossing the chemical potential, as described by our ARPES-base parameters for the Z point. The electron pocket is composed of two overlapping ellipses, roughly the same size as the hole pocket. As the temperature is increased, we observe a shrinking of the hole pocket and growth of the electron pocket, consistent with a rigid shift of the chemical potential as seen in other iron-based superconductors and described by our model. 
	
	By $\sim$200~K the inner hole band at the hole pocket shifts below the chemical potential, followed by the outer hole pocket by $\sim$300~K, suggesting multiple temperature induced Liftshitz transitions \cite{Wu2015}. The electron pocket grows monotonically and appears more circular by 300~K, which can be explained by the different Fermi velocities of the two overlapping electron bands, and has been observed in K-doped systems \cite{Miyata2015}.
	
	To determine the magnitude of the chemical potential shift we analyse the change of peak positions in the energy distribution curves (EDCs). In Fig.~4(a) and (b) we plot the EDCs for the centers of the hole and electron pocket respectively as a function of temperature. For the hole pocket, two main features are observed: the intense $d_{z^2}$ peak at $\sim220$~meV, and a complex peak close to the chemical potential which arises from the overlap of $d_{xz}$, $d_{yz}$ and $d_{xy}$ bands. For the electron pockets, the EDC has a single peak which originates from the slightly overlapping $d_{xz}/d_{yz}$ and $d_{xy}$ bands \cite{Watson2016}. It can be seen from both Fig.~4(a) and 4(b) that the  peak positions shift linearly as a function of temperature for all bands, further confirming a rigid band shift of the chemical potential. By extracting the peak position for the $d_{z^2}$ band in Fig.~4(a), which is composed of only a single feature and is situated far away from the chemical potential, we obtain a linear decrease of $\sim-0.13$~meV/K leading to a rigid band shift of $\sim$25~meV between 100~K and 300~K. 
	
	In Fig.~4(c) we plot the experimentally extracted $k_F$ values as a function of temperature for the hole pockets. Assuming that the chemical potential shift is proportional to temperature, and that the hole band dispersions are quadratic, we should expect to observe a square root behaviour of $k_F$ as a function of temperature. The square root functions fitted to the data (blue lines) show that the $k_F$ values for both the inner and outer hole bands are in agreement with this assumption and suggest that the entire hole pocket has shifted below the chemical potential by $\sim$300~K. We also present the $k_F$ values of the electron pockets as a function of temperature in Fig.~4(d). For the electron pockets, the band dispersions in the vicinity of the chemical potential are approximately linear as seen in Fig.~1. As a result the increase in the chemical potential leads to linear increases in $k_F$ values. The different Fermi velocities of the two overlapping electron pockets are also noticeable by the different slopes of the linear fits to the $k_F$ values. All the experimental results points towards a rigid increase of the chemical potential across the entire Brillioun zone.

	\begin{figure*}
		\includegraphics[width = \textwidth]{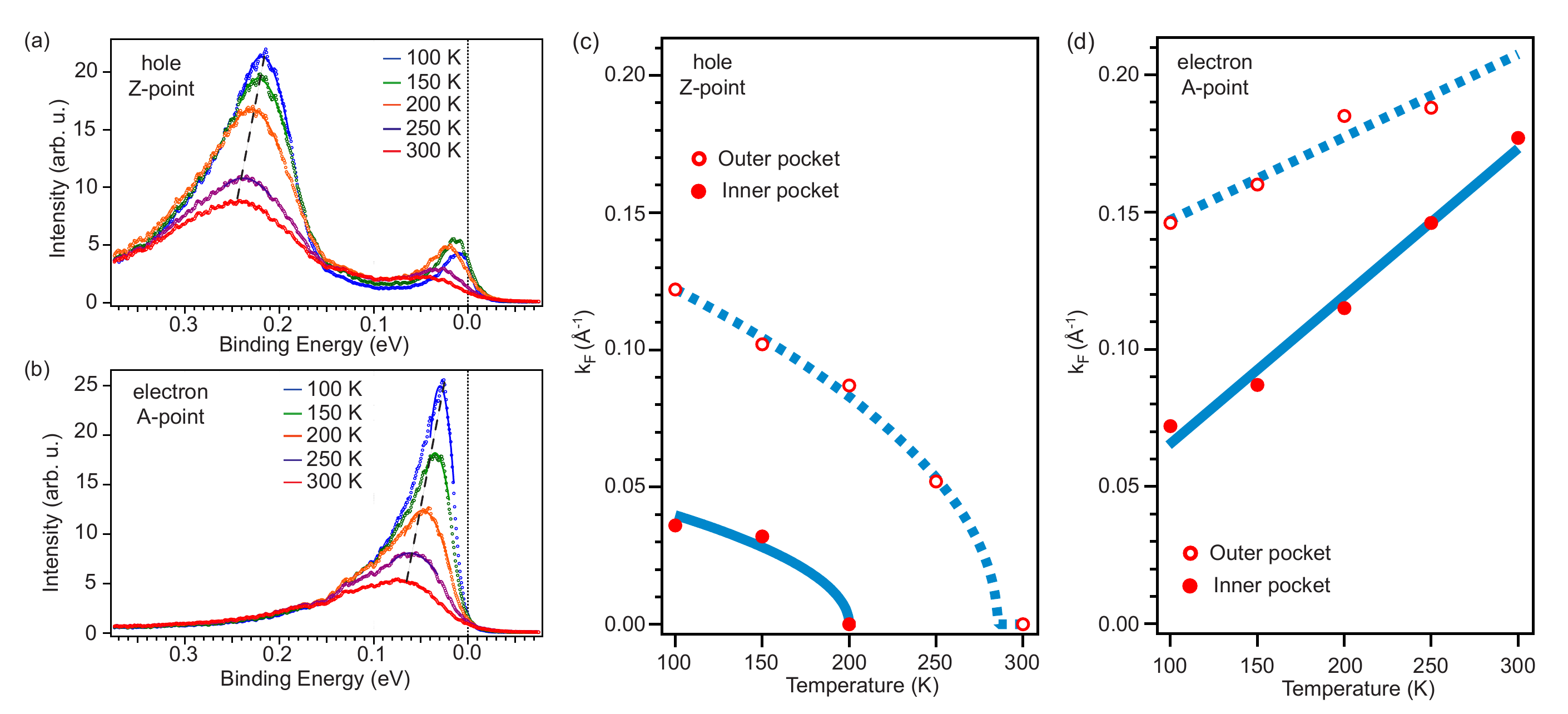}
		\caption{\label{figure4}  Energy distribution curves for the center of the hole (a) and electron (b) pockets as a function of temperature. (c,d) Measured $k_F$ values extracted from fit to momentum distribution curves at the Fermi level as a function of temperature, for the hole and electron point respectively. A square root fit was used for the trendline of (c) whilst a linear fit was used for (d).}
	\end{figure*}
	
	\section{Discussion}
	Whilst the ARPES-based parameter set predicts a significant temperature dependence of the chemical potential between 100~K and 300~K  (10~meV), the experimentally observed shift appears to be over twice as large (25~meV). Although there could be some limitations to the precision of our model, the most important features that affect the chemical potential shift, namely the position of the top of the hole band and bottom of the electron band with respect to the chemical potential, are very accurate to the experimental quasiparticle dispersions at 100~K. It does not seem likely that the enhancement of the chemical potential shift by a factor of two can be attributed to any numerical inaccuracy of the model. 
	
	We are thus led to the conclusion that
	this discrepancy must be attributed to temperature dependent electronic correlations, 
	which are not accounted for in our tight binding model. 
	Such correlations could lead to a reduction in the quasiparticle coherent weight with increasing temperature, as well as temperature-dependent life time effects and possibly to asymmetric spectral lineshapes.
	In particular, a decrease of quasiparticle coherent weight with increasing temperature
	would lead to a decrease in the density of states, which would require a greater increase to the chemical potential to compensate for the extra loss of states. 
	Such effects are often described in terms of self-energy effects, and have been discussed previously in theoretical studies of the iron pnictides.\cite{Heimes11, Heimes12}
	
	The unexpectedly large temperature dependence of the chemical potential in the range of $100-300$~K has important implications on the understanding of various physical properties of FeSe, particularly transport measurements and the spin fluctuation spectrum as probed by inelastic neutron scattering. Specifically the chemical potential shift at high temperatures will result in worse Fermi surface nesting properties between the hole and electron pockets, which could be relevant to the  suppression of ($\pi$,0) magnetic fluctuations at temperatures above 100 K \cite{Wang2016}. The chemical potential will also play a role at the nematic transition at 90 K, since any orbital order parameter which creates band shifts around the Fermi level will necessitate a shift of the chemical potential to preserve the total charge of the system. Thus the recently-observed momentum-independent downward shift of the $d_{xy}$ band observed below 90 K \cite{Watson2016,Watson2017} could be interpreted as being primarily a chemical potential effect. 
	
	\section{Summary}
	To conclude, we have determined a new set of tight-binding parameters 
	for the high temperature tetragonal phase of FeSe that is in good quantitative agreement with high resolution ARPES data at 100~K. 
	We have used these parameters to predict a large temperature dependence of the chemical potential in FeSe, that is not captured from DFT-based parameters. 
	We have performed a temperature-dependent ARPES study of FeSe up to room temperature and observed an enhanced 25~meV shift of the chemical potential between 100~K and 300~K. Our study provides a quantitatively accurate description of the low energy quasiparticle dispersions and highlights the necessity for full consideration of the chemical potential in future studies of FeSe.

	\section{Acknowledgments}
	We would like to thank A. I. Coldea, S. V. Borisenko, V. Sacksteder, M. Hoesch and R. Curtis for useful discussions. We thank Diamond Light Source for access to Beamline I05 (Proposals No. SI-12799, CM14454-5) that contributed to the results presented here. L. C. R. is supported by an iCASE studentship of the UK Engineering and Physical Sciences Research Council (EPSRC) and Diamond Light Source Ltd CASE award.
	
	\appendix
	\section*{Appendix}
	\subsection*{Tight binding optimization}
	Starting form the renormalized tight binding parameters from Ref.~[\onlinecite{Mukherjee2015}] we optimize the 28 2D and 17 3D tight binding parameters to the experimentally extracted band dispersions by a least squares method using the Powell optimization algorithm.\cite{Powell}. 
	Following the notation of Ref. [\onlinecite{Eschrig2009}], we label the hopping parameters as $t^{xy}_{\alpha\beta}$ where $x$ and $y$ describe the real space Fe-Fe hopping in the $x$ and $y$ direction of the quazi 2D plane, and $\alpha,\beta$ state the initial and final orbital the hopping parameter is describing. $\epsilon_\alpha$ describes the onsite energy of orbital $\alpha$  The orbitals are numbered as [$1:d_{xy}^+, 2:d_{x^2-y^2}^+, 3: id_{xz}^+,4:id_{yz}^+,5:d_{z^2}^+,6:d_{xy}^-, 7:d_{x^2-y^2}^-, 8:-id_{xz}^-,9:-id_{yz}^-,10:d_{z^2}^-$] where the + and - indicate if the orbital is on Fe site 1 or 2  . $\mu(100 K)$ is the chemical potential at 100~K. and $\lambda_{SOC}$ is the magnitude of the spin orbit coupling used in the model.
	Whilst most parameters are only altered by less than $\pm$30\% there are significant changes (over 100\%) in $t^{20}_{11}$, $t^{21}_{16}$,$t^{21}_{49}$ and $\epsilon_1$. 
	
	The hopping parameters obtained from the optimization are as follows:
	
	\begin{figure*}[t]
		\includegraphics[width = 0.89\textwidth]{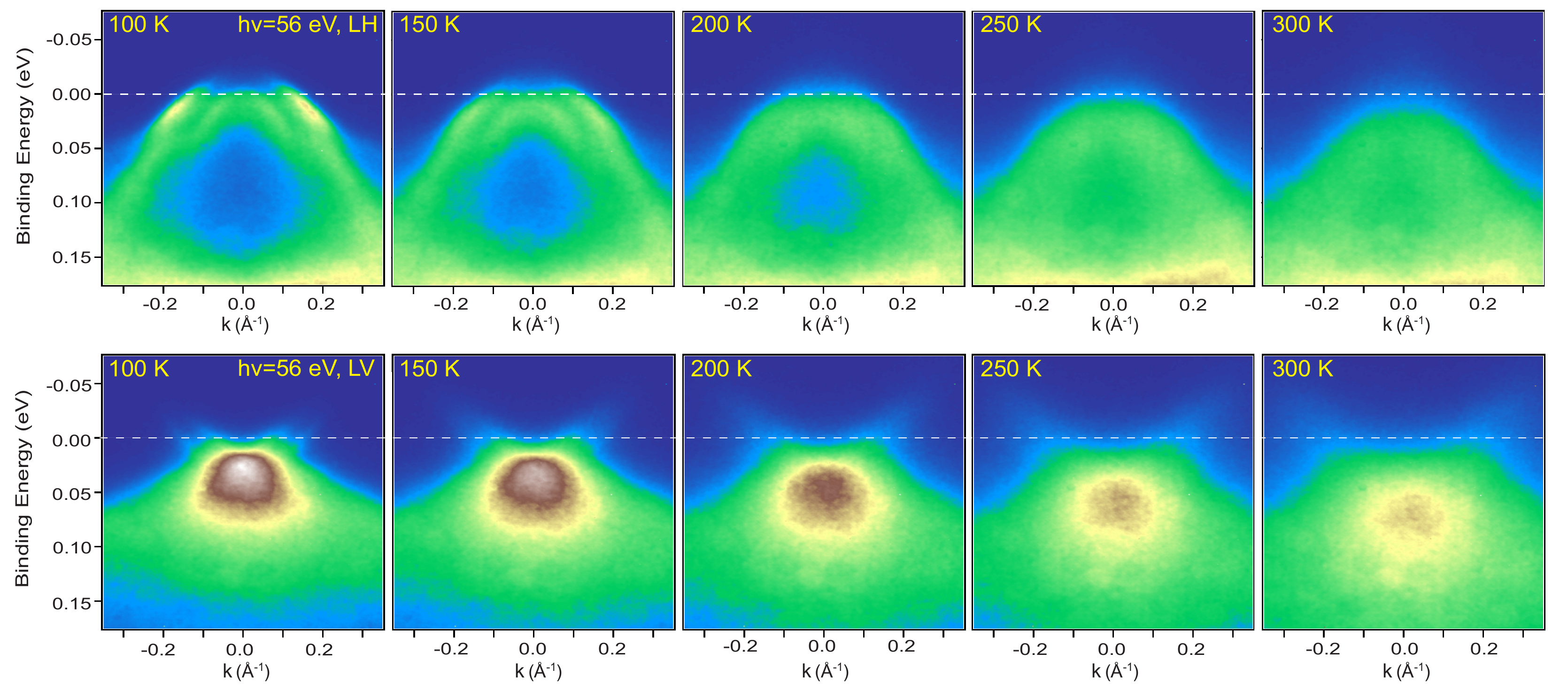}
		\caption{\label{figure5} Temperature dependent high symmetry band dispersions of the hole (top row) and electron (bottom row) pockets, taken with a photon energy of 56~eV corresponding to the Fermi surface maps in Fig. 3. }
	\end{figure*}

	\begin{align*}
		&\text{2D parameters}\nonumber \\
		t^{11}_{11}  &=  0.01818 
		&& t^{10}_{16}  = -0.03133
		&&t^{11}_{33}  =  0.02433 
		\nonumber \\
		t^{20}_{11}  &=  0.00093 
		&&t^{21}_{16}  = -0.00231
		&&t^{20}_{33}  = 0.00096 
		\nonumber \\
		t^{11}_{13}  &=-0.01226i
		&&t^{10}_{18}  =  0.11516i
		&&t^{02}_{33}  = -0.00717 
		\nonumber \\
		t^{11}_{15}  &= -0.01817
		&&t^{10}_{27}  = -0.04988
		&&t^{22}_{33}  =  0.00758 
		\nonumber \\
		t^{11}_{22}  &= -0.01669 
		&&t^{10}_{29}  = -0.09492i
		&&t^{10}_{38}  =  0.00868
		\nonumber \\
		t^{11}_{23}  &=  0.01484i 
		&&t^{10}_{2,10} =  0.059659
		&&t^{21}_{38}  = -0.00493
		\nonumber \\
		t^{11}_{34}  &=  0.01650 
		&&t^{11}_{35}  =  0.00569i 
		&&t^{10}_{4,10} =  -0.00902i
		\nonumber \\
		t^{10}_{49}  &=  0.05023
		&&t^{21}_{49} = -0.00008
		\nonumber \\
		\epsilon_1     &=  0.03405
		&&\epsilon_2     = -0.05050 
		&&\epsilon_3     =  0.00310
		\nonumber \\
		\epsilon_4    &=  0.00310 
		&&\epsilon_5     = -0.19398 \nonumber \\
		\mu(&100K) = 0.012 &&\lambda_{SOC} = 0.016 \nonumber
	\end{align*}

	\begin{align*}
		&\text{3D parameters} \nonumber \\
		t^{101}_{16} &= 0.00270
		&&t^{001}_{11} = 0
		&&t^{201}_{14} = 0.00950i
		\nonumber \\
		t^{121}_{16} &= -0.00283
		&&t^{111}_{11} = 0
		&&t^{101}_{19} = 0.00333i
		\nonumber \\
		t^{101}_{18} &= 0.00150i
		&&t^{201}_{11} =  0.00013
		&&t^{121}_{19} = 0.00517i
		\nonumber \\
		t^{001}_{33} &= 0.00183
		&&t^{101}_{38} = 0.00300
		&&t^{121}_{49} = 0.00100.
		\nonumber \\
		t^{201}_{33} &= -0.00133
		&&t^{121}_{38} = -0.00050
		&&t^{101}_{49} = 0.00217
		\nonumber \\
		t^{021}_{33} &= 0.00333
		&&t^{101}_{39} = 0.00250 
		\nonumber \\
	\end{align*}

	\subsection*{Band Dispersions}
	In Fig.~5 we present the band dispersions corresponding to the high symmetry cuts of Fig.~3. The white dashed line is the chemical potential position. A rigid shift of both the hole and electrons bands is observed, consistent with an increase of the chemical potential.

\end{document}